# Simple Mechanisms for Agents with Complements

Michal Feldman[*]    Ophir Friedler[†]    Jamie Morgenstern[‡]    Guy Reiner[§]

September 3, 2016

**Abstract**

We study the efficiency of simple auctions in the presence of complements. Devanur et al. [11] introduced the single-bid auction, and showed that it has a price of anarchy (PoA) of $O(\log m)$ for complement-free (i.e., subadditive) valuations. Prior to our work, no non-trivial upper bound on the PoA of single bid auctions was known for valuations exhibiting complements. We introduce a hierarchy over valuations, where levels of the hierarchy correspond to the degree of complementarity, and the PoA of the single bid auction degrades gracefully with the level of the hierarchy. This hierarchy is a refinement of the Maximum over Positive Hypergraphs (MPH) hierarchy [16], where the degree of complementarity $d$ is captured by the maximum number of neighbors of a node in the positive hypergraph representation. We show that the price of anarchy of the single bid auction for valuations of level $d$ of the hierarchy is $O(d^2 \log(m/d))$, where $m$ is the number of items. We also establish an improved upper bound of $O(d \log m)$ for a subclass where every hyperedge in the positive hypergraph representation is of size at most 2 (but the degree is still $d$). Finally, we show that randomizing between the single bid auction and the grand bundle auction has a price of anarchy of at most $O(\sqrt{m})$ for general valuations. All of our results are derived via the smoothness framework, thus extend to coarse-correlated equilibria and to Bayes Nash equilibria.

## 1 Introduction

A central focus of algorithmic mechanism design is to decide how to allocate limited resources to strategic agents while taking into account computational limitations. A great deal of work has studied truthful mechanisms, and while many times achieving guarantees that match the algorithmic problem (in which the agents are not strategic but always truth telling), many of the designed mechanisms turned out quite complex algorithmically and complicated to describe.

Practical concerns have led recent study to forgo truthfulness in lieu of simple mechanism formats. Simultaneous item auctions (SIAs), in particular, have constant-factor welfare approximations at equilibrium for subadditive buyers [18], and have an arguably simple format: each buyer submits a single sealed bid for each item separately, and each item's winner is the highest bidder for that item. Unfortunately, SIAs have a marked *lack* of simplicity in another respect: there is initial evidence that the problem of computing Nash equilibria [14], approximate Bayes Nash equilibria, correlated equilibria, or verifying best-responses [4] are likely intractable.

So, while SIAs have a simple format, the strategic behavior induced by the mechanism is quite complex. A mechanism with a simple format but one that is difficult to play leads one to question the underlying assumption that an equilibrium will be reached, and in turn to question the applicability of the price of anarchy bounds.

---

[*]Tel Aviv University; `michal.feldman@cs.tau.ac.il`; This work was partially supported by the European Research Council under the European Union's Seventh Framework Programme (FP7/2007-2013) / ERC grant agreement number 337122.

[†]Tel Aviv University; `ophirfriedler@gmail.com`

[‡]University of Pennsylvania; `jamiemmt.cs@gmail.com`

[§]Tel Aviv University; `guy.reiner@gmail.com`

Recent work [11] introduced another mechanism whose format was "simple" with a strategy space small enough that no-regret learning algorithms (for computing correlated and coarse correlated equilibria of the mechanism) run in polynomial time. This mechanism was coined the *single bid* mechanism, and was shown to have a Price of Anarchy (PoA) of $O(\log m)$ for subadditive buyers, where $m$ is the number of items. This upper bound on the PoA, while worse than that of SIAs, should apply to the welfare achieved by polynomially bounded agents (unlike those for SIAs).

The format of the single-bid mechanism was generalized by Braverman et al. [3], who defined the notion of a priori learnable interpolation (ALI) mechanisms. An ALI mechanism has two phases. First, agents report $O(\log m)$ bits of information to the mechanism. The mechanism computes some truthful mechanism as a function of all agents' reports. Second, the agents interact with this truthful mechanism. Since the second interaction is with a truthful mechanism, agents strategize only over their reports in the first phase. To find reports for the first round which form an equilibrium, one can trivially employ no-regret learning in polynomial time over the possible poly(m) reports. Thus, these mechanisms are strategically simple. If the truthful mechanism selected at the second phase always has a simple format, then the ALI mechanism will also have a simple format.

Both SIAs and single-bid auctions provide good approximation guarantees for complement-free (i.e., subadditive) bidders. However, valuations with *complementarities* arise naturally in many contexts, such as radio spectrum auctions, auctions for landing and takeoff time slots in airports, auctions for computational resources in the cloud, and more (see [8]).

In this work, we aim to design mechanisms for bidders with complementarities, which simultaneously approximate optimal welfare at equilibrium, have a simple format, and are strategically simple (as defined implicitly by Devanur et al. [11] and formally by Braverman et al. [3]). Formally, we wish to find mechanisms that run in polynomial time, whose equilibria have high welfare, and whose equilibria can be found in a computationally efficient manner, when bidders' valuations are not necessarily subadditive.

Several classes of valuations with restricted complements have been proposed in the literature: (1) positive hypergraphs with rank at most $k$ (PH-$k$), where the valuation is represented by a weighted hypergraph, the hyperedges have positive weights, and are of size at most $k$. The valuation for a set of items $S$ is the sum of the weights of the hyperedges contained in $S$. (2) maximum over PH-$k$ (MPH-$k$), where the value for a set of items $S$ is the maximum value assigned to $S$ across multiple PH-$k$ valuations. (3) supermodular-$d$ (SM-$d$), where the following graph is considered: the nodes correspond to goods, and an edge $(i, j)$ indicates complementarity between the goods $i$ and $j$[1]. The complementarity level $d$ corresponds to the maximum degree of any node in the graph.

While for SIAs, the PoA for MPH-$k$ is bounded by $2k$ [16], for single bid auctions, the PoA can be linear in $m$ even for PH-2.

Our main result is that a refinement of the MPH-$k$ hierarchy captures the degradation of the single bid auction in the presence of complements. This refinement is described next. The number of neighbors of a node in the PH representation of a valuation is the number of nodes with which a node shares a hyperedge. Our hierarchy, called *Maximum over Positive Supermodular d* (MPS-d), is a maximum over a collection of PH valuations in which the maximum number of neighbors of any node is bounded by $d$. Clearly, this constraint implies that each valuation in the collection is also in PH-$d$. One can verify that this refinement of PH-$d$ is equivalent to the intersection of PH and SM-$d$. This hierarchy is complete; the highest level of the hierarchy, MPS-m, captures all monotone functions. We show that the price of anarchy of the single bid auction degrades gracefully with the degree of the MPS-d hierarchy. This is cast in the following theorem.

**Theorem:** When agents have MPS-d valuations, the single-bid auction has a price of anarchy of at most $\frac{(d+1)}{1-e^{-(d+1)}} \cdot (d+2) \cdot H_{\frac{m}{d+1}} (= O(d^2 \log(m/d)))$ w.r.t. coarse correlated equilibria[2].

We also show that for a subclass of MPS-d valuations, where for every PH representation in the collection every hyperedge is of size at most 2 (i.e., where weights are given only to nodes and edges in the graph), and the number of neighbors of a node is bounded by $d$ (i.e., the intersection of PH-2 and SM-$d$), the price of anarchy is at most $O(d \log m)$.

---

[1]Goods $i$ and $j$ are said to exhibit complementarity if there exists some set $S$ such that $v(j|S \cup i) > v(j|S)$.
[2]For ease of exposition $H_x$ denotes the $x$-th harmonic number when $x$ is an integer and $H_{\lfloor x \rfloor} + 1$ otherwise.

We prove this by defining an extension of the constrained homogeneous class [11], called $d$-constrained homogeneous ($d$-CH), which allows for complements. Our proof proceeds in two steps. We first show that the $d$-CH class pointwise approximates MPS-$d$ valuations. We then show that the price of anarchy of the single bid auction with $d$-CH valuations is bounded by $O(d^2 \log(m/d))$.

The above results imply a good price of anarchy for valuations that lie in low levels of the MPS-d hierarchy. We then address the problem of simple auctions for general valuations. We show that randomizing between the single bid auction and the *grand bundle auction* (where the grand bundle is sold via a first price auction) obtains a price of anarchy of at most $O(\sqrt{m})$ for general valuations. Notably, running the two auctions in parallel (by soliciting independent bids) and choosing the better outcome of the two results in a price of anarchy of $\Omega(m)$. The desired result is obtained by randomizing between the two auction formats.

**Theorem:** The mechanism that randomizes between the single bid auction and the grand bundle auction achieves a price of anarchy of at most $\frac{4\sqrt{m}}{1-e^{-1}}$ for general valuations.

This bound matches the best poly-time welfare approximation by truthful mechanisms (assuming access to a demand oracle) [13, 15, 22, 28]. It is also known that SIAs cannot achieve a better price of anarchy bound for general valuations [21]. Finally, it should be noted that there exists a deterministic mechanism that has a price of anarchy of at most $O(\sqrt{m})$ for general valuations [24].

## 1.1 Related work

There has been a great deal of recent focus on simple mechanism design. These mechanisms achieve simplicity of format while trading off the optimality of the allocation they produce; the efficiency of simple, non-truthful mechanisms is measured using the price of anarchy. The goal of this line of research has been to design simple mechanisms whose price of anarchy is as small as possible in as general a setting as possible.

Sequential first-price item auctions have been shown to yield a constant price of anarchy for unit-demand bidders, with respect to subgame perfect equilibrium[3] [23] and Bayes-Nash equilibria [31]. This efficiency breaks for more general classes of valuations than unit-demand bidders: even with one additive bidder and $n-1$ unit-demand bidder, the pure Nash PoA can be $\Omega(m)$ [19].

The techniques for upper-bounding the Bayes-Nash PoA were shown to be generally useful: if one bounds a mechanism's PoA using a *smoothness* argument (introduced for auctions by Syrgkanis and Tardos [32], which is closely related to the smoothness of a game [29]), then PoA guarantees naturally extend to coarse correlated equilibria of the complete information game as well as Bayes-Nash equilibria.

The study of simultaneous item auctions was initiated by Christodoulou et al. [7], who showed that when buyers' valuations are submodular and i.i.d., the Bayesian PoA of second-price SIAs is at most 2, and that Pure Nash equilibria can be computed in polynomial time in the full-information setting for submodular buyers.

First-price simultaneous item auctions have been studied by Hassidim et al. [21]. They showed that pure Nash equilibria (when they exist) are fully efficient, but that mixed equilibria can have PoA of $\Omega(\sqrt{m})$ for general valuations. In addition, they showed that the price of anarchy for both coarse correlated equilibria with complete information and Bayes-Nash equilibria is $O(m)$ for general valuations, $O(\log m)$ for subadditive valuations, and $O(1)$ for XOS valuations.

SIAs were then shown by Feldman et al. [18] to have constant PoA at Bayes-Nash equilibria for subadditive buyers, for both first and second price payment rules. This result is tempered somewhat by a string of evidence suggesting that the problem of computing Nash equilibria [14] (for subadditive bidders), approximate Bayes-Nash equilibria (even for a mix of unit-demand and additive bidders), correlated equilibria, or verifying best-responses [4] are likely intractable.

---
[3]The natural extension of Nash Equilibrium to sequential games.

Another simple auction format that does allow for efficient computation of its coarse correlated equilibria (using no-regret learning algorithms and demand oracles) is the single-bid auction. In this auction, each bidder submits a single real number, and buyers (in descending order of their bids) choose a bundle amongst the remaining items, paying their bid for each item. This auction format was introduced by Devanur et al. [11], where the authors showed its price of anarchy of $O(\log m)$ for coarse correlated equilibria with subadditive bidders. The computational efficiency relied on the mechanism having a single round of strategic play which has a small action space, followed by a round of truthful behavior where agents select a utility-maximizing bundle. Braverman et al. [3] showed that this was essentially the best welfare one could achieve using any interpolation protocol which first has a single round of strategic play over a small action space, followed by some non-adaptive posted price mechanism.

Lucier and Borodin [24] give a deterministic greedy mechanism that has $O(\sqrt{m})$ price of anarchy. There are also truthful mechanisms that give $O(\sqrt{m})$ approximation to welfare, and run in poly time given an access to a demand oracle [13, 15, 22].

Several notions of hierarchical restricted complements have been introduced in the literature. Abraham et al. [1] introduce positive hypergraph representations of valuations with rank at most $k$, PH-$k$, give $k$-approximation algorithms for welfare approximation and $O(\log^k m)$-approximate truthful mechanisms for this class (and show the algorithmic result is the best possible in polynomial time unless $P = NP$). Feige and Izsak [17] introduce the notion of supermodular degree (at most) $d$, SM-$d$. When valuations are in SM-$d$, they show APX-hardness of answering demand queries for SM-$d$ for $d \geq 3$, and construct two $(d + 2)$-approximation algorithms for welfare maximization. Feige et al. [16] introduce a complete hierarchy of monotone functions, the maximum over positive hypergraphs with rank at most $k$, MPH-$k$. They give a $(k + 1)$-approximation to welfare maximization for this class, and show that SIAs have a price of anarchy at most $2k$ when buyers' valuations are contained in MPH-$k$.

Simple auction design has also been studied in the context of revenue maximization, both in single-parameter [10, 12, 20, 27] and multiparameter [2, 5, 6, 30, 34] contexts.

**Organization of the paper** In section 2 we formally define the setting and introduce our refined hierarchy, MPS-$d$. In section 3 we prove our main results, namely upper and lower bounds on the price of anarchy of the single bid auction with MPS-$d$ valuations. In section 4 we prove that randomizing between the single bid auction and the grand bundle auction achieves a price of anarchy of $O(\sqrt{m})$ for general valuations. We conclude in Section 5 with a discussion and some open problems.

## 2 Preliminaries

A combinatorial auction design problem consists of a set $N$ of $n$ agents, and a set of goods $[m] = \{1, 2, \ldots, m\}$. Each agent $i$ has a private valuation function $v_i : 2^{[m]} \to \mathbb{R}_+$. We use $\mathbf{v}$ to denote the valuation profile $(v_i)_{i \in N}$. We also write $\mathbf{v} = (v_i, \mathbf{v}_{-i})$, where $\mathbf{v}_{-i}$ denotes the valuations of all agents other than $i$. We design auctions which allocate each agent $i$ a set of goods $S_i$, such that the social welfare $\text{SW}(S) = \sum_i v_i(S_i)$ is (approximately) maximized. Let $\text{OPT}(\mathbf{v})$ be an allocation that maximizes the social welfare for the valuation profile $\mathbf{v}$. Fixing an auction and the behavior of all $n$ agents, each agent is charged some payment $P_i \geq 0$. An agent $i$ with valuation $v_i$ who is allocated a set of items $S_i$ and charged $P_i$ has quasi-linear utility $u_i = v_i(S) - P_i$. We will assume agents will behave to maximize this utility.

A mechanism is *truthful* if truth-telling is a dominant strategy; i.e., each agent maximizes its utility by reporting truthfully, regardless of its valuation and other agents' actions. An *interpolation mechanism* is a communication protocol with two phases. The first phase is non-truthful, and its output is a truthful mechanism.

**Definition 1** *(Braverman et al. [3]) An interpolation mechanism is* a priori learnable *if the first phase contains a single simultaneous broadcast round of communication, and the per-agent communication is $O(\log m)$.*

The following observation describes the key property that motivates the study of a priory learnable interpolation (ALI) mechanisms.

**Observation 2.1** *[3] An agent can run a regret-minimizing algorithm over her strategies in an a priori learnable interpolation mechanism (ALI) in time/space poly(m). Therefore, a correlated equilibrium of any ALI can be found in poly-time, and correlated equilibria arise as the result of poly-time distributed regret minimization.*

**The Single-bid auction** The single-bid auction, recently introduced by Devanur et al. [11], is an ALI mechanism. In the first phase the auctioneer solicits a single bid $b_i \in \mathbb{R}_+$ from each agent $i$. In the second phase the auctioneer sequentially approaches the agents, in a decreasing order of their bids (ties are broken arbitrarily), and offers each agent $i$ to purchase any of the items that have not been purchased yet, at a per-item price of $b_i$. We assume that agents maximize their utility: when offered a set of items $U \subseteq [m]$, agent $i$ selects a set $S_i \in \arg\max_{S \subseteq U} v_i(S) - |S_i| \cdot b_i$. Notice that fixing the first phase of the single-bid auction, the second phase is truthful; that is, reporting a set in $\arg\max_{S \subseteq U}\{v_i(S_i) - |S_i| \cdot b_i\}$ maximizes utility. Therefore, we assume that agent $i$ behaves strategically only when reporting her bid in the first phase, and truthfully selects a utility-maximizing set in the second phase. Assuming that a single bid can be expressed using communication size of $O(\log m)$, the singe bid auction is an ALI mechanism.

**Price of Anarchy and smoothness.** The allocation resulting from strategic play in the single-bid auction can result in a sub-optimal allocation of goods. Observation 2.1 implies that agents employing no-regret algorithms will converge to an (approximate) correlated or coarse correlated equilibrium. Therefore, it is of interest to provide efficiency guarantees on correlated and coarse equilibria. This efficiency is measured via the *price of anarchy* (PoA), which is the ratio of the optimal social welfare to the welfare at the worst possible equilibrium. Given an equilibrium $eq$, denote by SW($eq$) the social welfare at this equilibrium.

**Definition 2** *Let $E$ denote any solution concept for mechanism $\mathcal{M}$, and let $\mathcal{V}$ be a class of valuation profiles. Then the price of anarchy (PoA) and the price of stability (PoS) of $\mathcal{M}$ with respect to $E$ when the agents' valuation profile is in $\mathcal{V}$ are:*

$$PoA = \max_{\mathbf{v} \in \mathcal{V}} \max_{eq \in E} \frac{SW(OPT(\mathbf{v}))}{SW(eq)} \qquad PoS = \max_{\mathbf{v} \in \mathcal{V}} \min_{eq \in E} \frac{SW(OPT(\mathbf{v}))}{SW(eq)}$$

All our positive results apply to coarse correlated equilibria and Bayes-Nash equilibria.

**Definition 3** *(Coarse Correlated Equilibrium) An $\alpha$-coarse correlated equilibrium is a joint distribution $\sigma$ over bid vectors, such that for each agent $i$ and bid $b_i'$:*

$$\mathbb{E}_{\mathbf{b} \sim \sigma}[u_i(\mathbf{b})] \geq \mathbb{E}_{\mathbf{b} \sim \sigma}[u_i(b_i', \mathbf{b}_{-i})] - \alpha$$

The smoothness notion was introduced by Roughgarden [29] for general games, and Syrgkanis and Tardos [32] applied it to mechanisms. The smoothness framework provides a method for proving price of anarchy upper bounds for various solution concepts.

**Definition 4** *(Syrgkanis and Tardos [32]) A mechanism $\mathcal{M}$ is $(\lambda, \mu)$-smooth for a class of valuations $\mathcal{V} = \times_i \mathcal{V}_i$ if for any valuation profile $\mathbf{v} \in \mathcal{V}$, there exists a (possibly randomized) action profile $a_i^*(\mathbf{v})$ such that for every action profile $\mathbf{a}$:*

$$\sum_i \mathbb{E}_{a_i' \sim a_i^*(\mathbf{v})}[u_i(a_i', \mathbf{a}_{-i}; v_i)] \geq \lambda \cdot SW(OPT(\mathbf{v})) - \mu \sum_i P_i(\mathbf{a}) \tag{1}$$

**Theorem 2.2** *(Syrgkanis and Tardos [32]) If a mechanism is $(\lambda, \mu)$-smooth then the price of anarchy w.r.t. coarse correlated equilibria is at most $\frac{\max\{1,\mu\}}{\lambda}$.*

## 2.1 Categories of valuation functions

A set function $f : 2^{[m]} \to \mathbb{R}_+$ is *normalized* if $f(\emptyset) = 0$ and monotone if $f(T) \leq f(S)$ for every $T \subseteq S$. As standard, we assume that all valuations are normalized and monotone.

A *hypergraph representation* of a set function $f$ is a (normalized, but not necessarily monotone) set function $h$ such that for every set $S \subseteq [m]$ it holds that $f(S) = \sum_{T \subseteq S} h(T)$. One can easily verify that every set function $f$ has a unique hypergraph representation $h$.

A set function is complement-free, or subadditive, if for all $S, T \subseteq [m]$ it holds that $f(S \cup T) \leq f(S) + f(T)$.

When studying a class of valuations $\mathcal{V}$, it can be useful to also study the class $\max(\mathcal{V})$, as defined below.

**Definition 5** *Given a class of valuations $\mathcal{V}$, the class $\max(\mathcal{V})$ is the class of all valuations that can be represented as a maximum over a collection of valuations from $\mathcal{V}$, i.e., $\max(\mathcal{V}) = \{f : \exists \mathcal{G} \subseteq \mathcal{V} : \forall S \subseteq [m], f(S) = \max_{g \in \mathcal{G}} g(S)\}$.*

In this paper we focus on valuation functions that exhibit complements. The following hierarchies of valuations with complements have been considered in the literature.

**Maximum over positive hypergraphs** [16] The class PH (*positive-hypergraph*) is the class of all functions $f$ whose hypergraph representation $h$ has nonnegative edges. The class PH-$k$ contains all functions $f \in$ PH for which every set $T$ with $h(T) > 0$ satisfies $|T| \leq k$. The class *maximum over PH-$k$* (MPH-$k$) is the class $\max(\text{PH-}k)$. Unlike PH-$k$, MPH-$k$ is a *complete hierarchy*: for every set function $f$, there exists some $k \leq m$ such that $f$ is in MPH-$k$ (in particular, all functions are in MPH-$m$).

**The supermodular degree** [17] The *supermodular degree* measures the extent to which any set function $f$ exhibits supermodular behavior. For an item $j$ and set $S$, denote by $f(j|S) = f(S \cup j) - f(S)$[4] the marginal value of item $j$ given $S$. The supermodular dependency set of item $j$ is defined as $\text{Dep}^+(j) = \{j' : \exists S \subseteq [m] \text{ so that } f(j|S \cup j') > f(j|S)\}$. The supermodular degree of $f$ is defined as $\max_{j \in [m]} |\text{Dep}^+(j)|$. The class *supermodular degree $d$* (SM-$d$) contains all the set functions with supermodular degree at most $d$. Clearly, the SM-$d$ hierarchy is complete, as any set function has supermodular degree at most $m - 1$.

## 2.2 A refined hierarchy of restricted complements

The lowest level in the MPH-$k$ hierarchy (MPH-1) is contained in the class of subadditive valuations. It follows from Devanur et al. [11] that for MPH-1 valuations the price of anarchy is upper bounded by $\frac{e}{e-1} H_m$ (where $H_m$ is the $m$'th harmonic number). However, this positive result does not extend beyond the lowest level of MPH (or even PH).

**Observation 2.3** *[26] The single bid auction has price of stability of at least $m$ when agents have valuations in PH-2.*

*proof.* A $t$-star-graph, centered at $j$, is a graph with $t$ nodes, where there is an edge between the center node ($j$) and each one of the other $t - 1$ nodes. A *$t$-star-shaped valuation* is a valuation with a $t$-star-graph hypergraph representation, in which all edges have weight 1.

Consider two agents, $a$ and $b$, and the items $[m]$. Let $v_a$ be an $m$-star-shaped valuation, centered at item 1. Therefore, for all $T \subseteq [m]$, $v_a(T) = |T| - 1$ if $1 \in T$ and 0 otherwise. By construction, $v_a \in$ PH-2. Agent $b$ only wants item 1 for a value of $(m-1)/m + \epsilon$. For agent $a$ to purchase item 1 in equilibrium, it must pay at least $(m-1)/m + \epsilon$, otherwise, agent $b$ can bid slightly higher than $a$'s bid and improve its utility. However, if agent $a$ acquires a set $T \ni 1$ for a price $p$ per item, its utility is $|T| \cdot (1 - p) - 1$. Therefore, if agent $a$ bids more than $(m-1)/m$, buying any set of items yields negative utility. As a result, at any equilibrium, agent $b$ gets item 1, agent $a$ has 0 value, and the social welfare is $(m-1)/m+\epsilon$. In the optimal outcome, agent $a$ gets all the items and the social welfare is $m-1$. Therefore, the fraction of the optimal welfare that is achieved in any pure equilibrium is $\frac{(m-1)/m+\epsilon}{m-1} = 1/m + \frac{\epsilon}{m-1}$. □

---
[4] We abuse notation and write $S \cup j$ instead of $S \cup \{j\}$

This bound is essentially tight. Indeed, it is easy to show[5] that the single bid auction is $((1 - e^{-m})/m, 1)$-smooth for *general valuations*, implying a price of anarchy of at most $m/(1 - e^{-m})$. This example demonstrates that the second level of the MPH hierarchy contains valuations that render the worst possible setting for the single bid auction. While one may interpret this result to imply that single bid auctions are hopeless in the presence of complements, we show that viewing restricted complements using a different lens reveals the effect of the level of complementarity on the performance of the auction. In particular, we establish positive results for a refined hierarchy, which combines the structural properties of both SM-$d$ and MPH-$k$ valuations. One would hope that the SM-$d$ hierarchy, by itself, would enable positive price of anarchy results. This is left as an open problem of this work.

**Maximum over Positive-Supermodular-$d$** Positive Supermodular $d$ (PS-$d$) functions are functions that have a PH representation, and furthermore their supermodular degree is at most $d$. Thus, in PS-$d$ valuations each item can have at most $d$ items in its supermodular dependency set (in the sense of [17]).

**Definition 6** *(Maximum over Positive-Supermodular-d) The class Positive Supermodular (PS-d) is defined as PS-d = SM-d ∩ PH, and the class MPS-d is defined as MPS-d= max(PS-d).*

Lemma 2.4 shows that PS-$d$ can be equivalently defined as the class of valuations with PH representation in which the number of neighbors of every node is at most $d$. This trivially implies that PS-$d$ is contained in PH-$(d+1)$.

**Lemma 2.4** *Let $v$ be a valuation in PS-d with a hypergraph representation $w$. For any two items $j, j' \in [m]$, it holds that $j' \in Dep^+(j)$ if and only if there exists a hyperedge $e$ for which $w_e > 0$ and $\{j, j'\} \subseteq e$.*

*proof.* For an item $j \notin S$ it holds that $v(j|S) = \sum_{e \subseteq S \cup j} w_e - \sum_{e \subseteq S} w_e = \sum_{e \subseteq S \cup j : j \in e} w_e$, therefore, for two items $j' \neq j$ not in $S$ it holds that: $v(j|S \cup j') - v(j|S) = \sum_{e \subseteq S \cup \{j,j'\}: j \in e} w_e - \sum_{e \subseteq S \cup j : j \in e} w_e = \sum_{e \subseteq S \cup \{j,j'\}: \{j,j'\} \subseteq e} w_e$. Therefore $j' \in \text{Dep}^+(j)$ if and only if the last sum is positive for some $S \in [m] \setminus \{j, j'\}$, which in turn holds if and only if $w_e > 0$ for some $e$ so that $\{j, j'\} \subseteq e$. □

The MPS-d hierarchy is *complete*[6], i.e., for every monotone valuation $f$ there exists some $d \leq (m-1)$ such that $f \in$ MPS-d.

## 3 The Single Bid Auction In The Presence Of Complementarities

We now present the main result of this section, namely, that the welfare in any coarse correlated equilibrium of the single-bid auction, when buyers' valuations are in MPS-d, is an approximation to the optimal welfare.

**Theorem 3.1** *For agents with valuations in MPS-d, the coarse correlated price of anarchy of the single-bid auction is no more than $\frac{1}{1-e^{-(d+1)}}(d+1)(d+2) \cdot H_{\frac{m}{d+1}}$.*

Specifically, we show that when agents have MPS-d valuations, the single bid auction is a $\left(\frac{1-e^{-(d+1)}}{(d+1)\cdot(d+2)\cdot H_{\frac{m}{d+1}}}, 1\right)$-smooth mechanism.

In addition, we prove a stronger upper bound of $\frac{2(d+1)}{1-e^{-2}} \cdot H_{m/2}$ when agents have max(PH-2 ∩ SM-$d$) valuations, which is a strict subclass of MPS-d. We also show a PoS lower bound of $\Omega(d + \frac{\log m}{\log \log m})$ when agents have PH-2 ∩ SM-$d$ valuations.

---
[5] As a corollary from Lemma 4.4
[6] Since PS-(m-1) = SM-(m-1) ∩ PH = PH, we get that MPS-(m-1) = MPH-$m$.

The following proof method for establishing the smoothness of a mechanism with respect to a class of valuations $\mathcal{V}$ was presented in Devanur et al. [11]: first show smoothness for a restricted class of valuations $\mathcal{V}'$. Then, show that the class $\mathcal{V}$ can be *pointwise $\beta$-approximated* by the restricted class $\mathcal{V}'$. Pointwise approximation is defined as follows:

**Definition 7** *[11]* [pointwise $\beta$-approximation] *A valuation class $\mathcal{V}$ is pointwise $\beta$-approximated by a valuation class $\mathcal{V}'$ if for any valuation $v \in \mathcal{V}$ and for any set $S \subseteq [m]$, there exists a valuation $v' \in \mathcal{V}'$ such that $\beta \cdot v'(S) \geq v(S)$ and for all $T \subseteq [m]$ it holds that $v'(T) \leq v(T)$.*

Note that pointwise $\beta$-approximation is less restrictive than mapping each valuation $v \in \mathcal{V}$ to a single valuation $v' \in \mathcal{V}'$ that approximates it everywhere, yet smoothness of a mechanism for valuations in $\mathcal{V}'$ implies smoothness for the larger class $\mathcal{V}$.

**Lemma 3.2** *[11] If a mechanism for a combinatorial auction setting is $(\lambda, \mu)$-smooth for the class of valuations $\mathcal{V}'$ and $\mathcal{V}$ is pointwise $\beta$-approximated by $\mathcal{V}'$, then it is $\left(\frac{\lambda}{\beta}, \mu\right)$-smooth for the class $\mathcal{V}$.*

A constraint-homogeneous (CH) valuation is an additive valuation such that the value of every item is either 0 or $\hat{v}$ for some fixed $\hat{v} > 0$. In Devanur et al. [11] it was proved that complement-free valuations are pointwise $H_m$-approximated by CH valuations.

When trying to apply a similar technique for the case of PS-d valuations, we face a challenge, namely that for $d \geq 1$ PS-d valuations cannot be pointwise $\beta$-approximated by complement-free valuations for any $\beta$. To see this, consider an instance with two items $\{a, b\}$ and the PS-1 valuation $v(\{a\}) = v(\{b\}) = 0$ and $v(\{a, b\}) = 1$. Any complement-free valuation $v' \leq v$ will have $v'(\{a\}) = v'(\{b\}) = 0$ which implies $v'(\{a, b\}) = 0$. Therefore, in order to use the technique of pointwise approximation for PS-d valuations one must go beyond complement-free valuations. To this end we introduce the following class of valuations.

**Definition 8** *(d-Constraint Homogeneous Valuations) A valuation $v$ is d-constraint homogeneous (d-CH) if there exists a value $\hat{v}$, and disjoint sets of items $Q_1, \ldots, Q_\ell$, each of size at most $d$, so that $v(Q_i) = \hat{v} \cdot |Q_i|$ for every $Q_i$, and the value of every set $S \subseteq [m]$ is the sum of values of contained $Q_i$'s, i.e.,*

$$v(S) = \sum_{Q_i \subseteq S} v(Q_i) = \hat{v} \sum_{Q_i \subseteq S} |Q_i| = \hat{v} \cdot |\{t : \exists i \text{ s.t. } t \in Q_i \subseteq S\}|$$

Note that 1-CH valuations are CH valuations and that $d$-CH valuations contain single minded bidders where the interest set of each agent is of size at most $d$. The remainder of this section is structured as follows. In Lemma 3.4 we show that when agents have $d$-CH valuations the single bid auction is a $(\frac{1-e^{-d}}{d}, 1)$-smooth mechanism. In Lemma 3.5 we show that the class of PS-d valuations is pointwise $(d+2) \cdot H_{\frac{m}{d+1}}$-approximated by $(d+1)$-CH valuations. These two lemmas imply the smoothness result for PS-d. Finally, Observation 3.3[7] implies that the same smoothness result carries over to MPS-d.

**Observation 3.3** *For every valuation class $\mathcal{V}$, the valuation class $\max(\mathcal{V})$ is pointwise 1-approximated by $\mathcal{V}$.*

We begin by proving smoothness for agents with $d$-CH valuations.

**Lemma 3.4** *The single bid auction is a $((1 - e^{-d})/d, 1)$-smooth mechanism when agents have d-CH valuations.*

---

[7]Observation 3.3 appeared previously (e.g. Lucier and Syrgkanis [25], Syrgkanis and Tardos [32]) and its proof is by definition: for a valuation $v \in \max(\mathcal{V})$ and a set $S \subseteq [m]$, let $v^* = v_\ell$ so that $\ell \in \arg\max_{\ell \in \mathcal{L}} v_\ell(S)$, then by definition $v(S) = v^*(S)$ and $v(T) \geq v^*(T)$ For any set $T \subseteq [m]$.

*proof.* Fix a valuation profile $\mathbf{v}$ of $d$-CH valuations, and let $S^* = \text{OPT}(\mathbf{v})$ be an optimal allocation w.r.t. $\mathbf{v}$. Fix an agent $i$, let $\hat{v}$ and $\{Q_\ell\}_\ell$ be the parameters in agent $i$'s valuation, and for presentation clarity write $v = v_i$; $v(S) = \hat{v} \cdot \sum_{Q_\ell \subseteq S} |Q_\ell|$. Consider a bid profile $\mathbf{b}$ and denote by $p_j(\mathbf{b})$ the induced price for item $j$, i.e., $p_j(\mathbf{b}) = b_{i^*}$ so that $i^*$ is the agent that purchases $j$ under bid profile $\mathbf{b}$. Consider an arbitrary set $Q_\ell \subseteq S_i^*$. Agent $i$ can acquire all items in $Q_\ell$ by bidding $t > \max_{j \in Q_\ell} p_j(\mathbf{b})$. In such a case the utility from purchasing $Q_\ell$ is $v(Q_\ell) - t \cdot |Q_\ell| = \hat{v} \cdot |Q_\ell| - t \cdot |Q_\ell| = |Q_\ell| \cdot (\hat{v} - t)$ Therefore:

$$u_i(t, \mathbf{b}_{-i}) \geq \sum_{Q_\ell \subseteq S_i^*} |Q_\ell| \cdot (\hat{v} - t) \cdot \mathbf{1}\{t > \max_{j \in Q_\ell} p_j(\mathbf{b})\}$$

Suppose $i$ performs the randomized deviation $a_i^*(v_i)$ with the density function $f(t) = \frac{1}{d} \cdot \frac{1}{\hat{v}-t}$ and support $[0, (1-e^{-d}) \cdot \hat{v}]$, Then:

$$\mathbb{E}_{t \sim a_i^*(v_i)}[u_i(t, \mathbf{b}_{-i})] \geq \sum_{Q_\ell \subseteq S_i^*} |Q_\ell| \cdot \int_{\max_{j \in Q_\ell}\{p_j(\mathbf{b})\}}^{(1-e^{-d})\hat{v}} (\hat{v} - t) \cdot f(t) \mathrm{d}t$$

$$= \frac{1}{d} \cdot \sum_{Q_\ell \subseteq S_i^*} |Q_\ell| \cdot \left((1-e^{-d})\hat{v} - \max_{j \in Q_\ell}\{p_j(\mathbf{b})\}\right)$$

By $\max_{j \in Q_\ell}\{p_j(\mathbf{b})\} \leq \sum_{j \in Q_\ell} p_j(\mathbf{b})$ and $v(Q_\ell) = \hat{v} \cdot |Q_\ell|$ and $|Q_\ell| \leq d$ we get that:

$$\mathbb{E}_{t \sim a_i^*(v_i)}[u_i(t, \mathbf{b}_{-i})] \geq \frac{1-e^{-d}}{d} \cdot \sum_{Q_\ell \subseteq S_i^*} v(Q_\ell) - \sum_{Q_\ell \subseteq S_i^*} \sum_{j \in Q_\ell} p_j(\mathbf{b})$$

Finally, the first sum is exactly agent $i$'s valuations for $S_i^*$, and the second sum is at most $\sum_{j \in S_i^*} p_j(\mathbf{b})$ since $\{Q_\ell\}_\ell$ is a partition, therefore:

$$\mathbb{E}_{t \sim a_i^*(v_i)}[u_i(t, \mathbf{b}_{-i})] \geq \frac{1-e^{-d}}{d} \cdot v(S_i^*) - \sum_{j \in S_i^*} p_j(\mathbf{b})$$

Summing over all agents establishes the smoothness property.
□

Note that the class of single-minded bidders with interest sets of size at most $d$ is a special case of $d$-CH valuations, so Lemma 3.4 implies a corresponding bound on the PoA of SBA with respect to single-minded valuations as well.

Next we show that the class PS-$d$ can be pointwise $(d + 2) \cdot H_{\frac{m}{d+1}}$-approximated by $(d + 1)$-CH valuations[8]. In the proof, we use the following two properties of PS-$d$ valuations: First, two items are in the super-dependency set of each other if and only if they share a hyperedge with a positive weight. Second, the size of the super-dependency set of an item is bounded by the level of the hierarchy. We note that neither the class SM-$d$ nor the class PH-$k$ (for $k \geq 2$) exhibit both properties.

**Lemma 3.5** *The PS-$d$ valuation class is pointwise $(d+2) \cdot H_{\frac{m}{d+1}}$-approximated by the $(d+1)$-CH valuation class.*

*proof.* Consider a valuation $v \in$ PS-$d$, a set $X \subseteq [m]$ and some $\beta$ to be determined later. Let $w$ be the hypergraph representation of $v$, i.e., $v(S) = \sum_{T \subseteq S} w_T$. Consider the following greedy construction of a partition $\mathcal{Q} = \{Q_\ell\}_\ell$ of the set $X$: While there are more than $d + 1$ items, select a subset of yet unselected $d + 1$ items from $X$, with maximum value (with respect to $v$). The remaining items form the last subset of the partition. The formal description of the greedy process is given in Algorithm 1.

---
[8]Our proof method is in the spirit of the proof that subadditive valuations are pointwise $H_m$-approximated by CH valuations, as appears in Devanur et al. [11]

**ALGORITHM 1:** Algorithm 1: Partitioning of set $X$.

**Input:** A set $X \subseteq [m]$, access to a valuation function $v$.
**Output:** A partition $\mathcal{Q} = \{Q_\ell\}_\ell$ of $X$

1   $S \leftarrow X$.
2   **for** each $\ell$ from 1 to $\lceil \frac{m}{d+1} \rceil$ **do**
3     Select a set $Q_\ell$ in $\arg\max_{\substack{A \subseteq S \\ |A| = d+1}} \{v(A)\}$, or $Q_\ell := S$ if $|S| < d+1$.
4     $S \leftarrow S \setminus Q_\ell$. If $S = \emptyset$ then terminate.
5   **end**

Let $h_\mathcal{Q}$ be the function:

$$h_\mathcal{Q}(T) = \frac{v(X)}{|\bigcup_\ell Q_\ell|\beta} \cdot \sum_{Q_\ell \subseteq T} |Q_\ell|$$

Note that for any family of disjoint subsets $\mathcal{Q}'$ each of size at most $d+1$, $h_{\mathcal{Q}'}$ is a $(d+1)$-CH valuation. It suffices to find some $\mathcal{Q}' \subseteq \mathcal{Q}$ so that $\beta \cdot h_{\mathcal{Q}'}(X) \geq v(X)$ and also $h_{\mathcal{Q}'}(T) \leq v(T)$ for all $T \subseteq [m]$. We will examine a sequence of such functions $h_{\mathcal{Q}'}$, so that if none of them pointwise $\beta$-approximates $v$ at $X$, then this implies an upper bound on $\beta$.

Initially consider $S_1 = X$. Since $\mathcal{Q}$ is a partition of $S_1$ we have that $h_\mathcal{Q}(X) = \frac{v(X)}{|X|\beta} \cdot \sum_\ell |Q_\ell| = \frac{v(X)}{\beta}$, so the first requirement of pointwise $\beta$-approximation holds. If $h_\mathcal{Q}(T) \leq v(T)$ for all $T \subseteq [m]$ then $h_\mathcal{Q}$ pointwise approximates $v$ at $|X|$. Otherwise, there exists some $T_1$ so that $h_\mathcal{Q}(T_1) > v(T_1)$. Since $v$ is monotone $v(\cup_{Q_\ell \subseteq T_1} Q_\ell) \leq v(T_1) < h_\mathcal{Q}(T_1) = h_\mathcal{Q}(\cup_{Q_\ell \subseteq T_1} Q_\ell)$ therefore we may assume w.l.o.g. that $T_1$ is a union of sets from $\mathcal{Q}$. Iteratively, consider $S_i = S_{i-1} \setminus T_{i-1}$. Since $T_{i-1}$ and $S_{i-1}$ are each a union of sets from $\mathcal{Q}$, then $S_i$ is also a union of sets from $\mathcal{Q}$, and $\mathcal{Q}_{S_i} = \{Q_\ell \in \mathcal{Q} : Q_\ell \subseteq S_i\}$ is a partition of $S_i$. By definition, $h_{\mathcal{Q}_{S_i}}(T) = \frac{v(X)}{|S_i|\beta} \sum_{Q_\ell \in \mathcal{Q}_{S_i}: Q_\ell \subseteq T} |Q_\ell|$ is a $(d+1)$-CH valuation, and since $\mathcal{Q}_{S_i}$ is a partition of $S_i$ we get that $h_{\mathcal{Q}_{S_i}}(X) = \frac{v(X)}{\beta}$. If for some $i$ it holds that $h_{\mathcal{Q}_{S_i}}(T) \leq v(T)$ for all $T \subseteq [m]$, then $h_{\mathcal{Q}_{S_i}}$ pointwise $\beta$-approximates $v$ at $X$. Otherwise, at some point the iterative process terminates and we are left with two partitions of the set $X$: $\{Q_\ell\}_\ell$ and $\{T_i\}_i$, so that every $Q_\ell$ is a subset of some $T_j$. Therefore:

$$\sum_\ell v(Q_\ell) \leq \sum_i v(T_i) < \sum_i h_{\mathcal{Q}_{S_i}}(T_i) = \frac{v(X)}{\beta} \sum_i \frac{|T_i|}{|S_i|} \quad (2)$$

where the first inequality is because $v$ has a positive-hypergraph representation, the second inequality is by construction, and the last equality is because every $S_i$ and $T_i$ are unions of subsets from $\mathcal{Q}$. Denote by $\mathcal{C}(\mathcal{Q})$ the collection of all hyperedges $e \subseteq X$ with $w_e > 0$ so that $e \not\subseteq Q_\ell$ for all $\ell$. By construction it holds that $v(X) = \sum_\ell v(Q_\ell) + \sum_{e \in \mathcal{C}(\mathcal{Q})} w_e$. The first sum in the last expression is the total weight of all (hyper)edges that are in the interior of some partition element $\mathcal{Q}_\ell$. The second is the total weight of all edges that connect at least two partition elements. We establish the following lemma:

**Lemma 3.6** $\sum_{e \in \mathcal{C}(\mathcal{Q})} w_e \leq (d+1) \sum_\ell v(Q_\ell)$

Before proving Lemma 3.6 we show how it is used to conclude the proof. Note that the proof of Lemma 3.6 relies on the properties of the class PS-d. Lemma 3.6 implies $v(X) \leq (d+2) \sum_\ell v(Q_\ell)$. By equation (2) we get: $v(X) < (d+2) \frac{v(X)}{\beta} \sum \frac{|T_i|}{|S_i|}$ therefore $\beta < (d+2) \sum \frac{|T_i|}{|S_i|}$. For ease of exposition assume $|X|$ is divisible by $(d+1)$, which implies that the cardinality of every $Q_\ell$, and hence every $S_i$ and every $T_i$ are divisible by $d+1$. Let $s_i = \frac{|S_i|}{d+1}$ and $t_i = \frac{|T_i|}{d+1}$. Therefore:

$$\sum_i \frac{|T_i|}{|S_i|} = \sum_i \frac{t_i}{s_i} = \sum_i \sum_{j=0}^{t_i - 1} \frac{1}{s_i} \leq \sum_i \sum_{j=0}^{t_i - 1} \frac{1}{s_i - j} = \sum_{j=0}^{s_1 - 1} \frac{1}{s_1 - j} = H_{s_1} = H_{\frac{|X|}{d+1}} \quad (3)$$

Which concludes that $\beta < (d+2) \cdot H_{\frac{m}{d+1}}$. It remains to prove Lemma 3.6.

*proof.* For each $Q_\ell$, we show there exists a set $E_\ell \subseteq \mathcal{C}(\mathcal{Q})$, such that the collection $\{E_\ell\}_\ell$ satisfies $\mathcal{C}(\mathcal{Q}) \subseteq \cup_\ell E_\ell$, and for every $\ell$ it holds that:

$$\sum_{e \in E_\ell} w_e \leq (d+1)v(Q_\ell) \tag{4}$$

We conclude that $\sum_{e \in \mathcal{C}(\mathcal{Q})} w_e \leq \sum_\ell \sum_{e \in E_\ell} w_e \leq (d+1) \sum_\ell v(Q_\ell)$, where the first inequality is true since $\mathcal{C}(\mathcal{Q}) \subseteq \cup_\ell E_\ell$. Let $E_\ell$ denote the set of hyperedges $e \in \mathcal{C}(\mathcal{Q})$ such that $\ell$ is the minimal index of a set from the partition $\mathcal{Q}$ for which $e \cap Q_\ell \neq \emptyset$. For every item $j \in Q_\ell$ define $E_\ell^j = \{e \in E_\ell : j \in e\}$, i.e., the hyperedges in $E_\ell$ in which $j$ is a member, clearly $E_\ell = \bigcup_{j \in Q_\ell} E_\ell^j$. For a set of hyperedges $E$, let $V(E) = \bigcup_{e \in E} e$. By Lemma 2.4 we get that $V(E_\ell^j) \subseteq (\text{Dep}^+(j) \cup \{j\})$[9], which implies that $\left|V(E_\ell^j)\right| \leq |Dep^+(j)| + 1 \leq (d+1)$, where the last inequality follows from PS-$d \subseteq$ SM-$d$. By definition of $E_\ell$, for every $j' \in V(E_\ell)$, if $j' \in Q_{\ell'}$, then $\ell' \geq \ell$, which implies that prior to the $\ell^{th}$ iteration of step 3 in Algorithm 1, all the items in $V(E_\ell)$ are available, i.e., in the set $S$. Therefore, for every item $j \in Q_\ell$ the set $V(E_\ell^j)$ was available. By step 3 and monotonicity of $v$, $Q_\ell$ maximizes value over all available sets of size at most $d+1$ therefore $v(Q_\ell) \geq v(V(E_\ell^j))$ for every $j$. Therefore:

$$\sum_{e \in E_\ell} w_e \leq \sum_{j \in Q_\ell} \sum_{e \in E_\ell^j} w_e \leq \sum_{j \in Q_\ell} v\left(V(E_\ell^j)\right) \leq |Q_\ell| v(Q_\ell) \leq (d+1)v(Q_\ell)$$

□
□

## 3.1 Improved PoA when hyperedges are of size at most $2$

In the following subsection we show that PH-2∩SM-$d$ valuations are pointwise $(d+1)H_{m/2}$-approximated by 2-CH valuations, which by Lemma 3.4 and observation 3.3 implies that the PoA is at most $\frac{2(d+1)H_{m/2}}{1-e^{-2}}$ when agents have valuations in max(PH-2 ∩ SM-$d$):

**Theorem 3.7** *The single bid auction is a $(\frac{1-e^{-2}}{2(d+1)H_{m/2}}, 1)$-smooth mechanism for max(PH-2 ∩ SM-d) valuations. Thus, it has a price of anarchy of at most $O(d \log(m))$.*

This shows an improvement of roughly a factor of $d$ when compared to MPS-$d$ valuations. The main difference when comparing to the proof of Theorem 3.1 is that we show that $max$(PH-2∩SM-$d$) valuations are $(d+1)H_{m/2}$-pointwise approximated by 2-CH valuations (as opposed to $(d+1)$-CH valuations).

**Lemma 3.8** *The class PH-2 ∩ SM-d is pointwise $(d+1)H_{m/2}$-approximated by 2-CH valuations*

*proof.* Let $v \in$ PH-2 ∩ SM-$d$ be a valuation function, and let $X$ be a set of items. W.l.o.g. assume $X = [m]$, and both terms will be used interchangeably during the proof. Let $G = (V, E)$ be its graphical representation with weights $w_e \geq 0$ for edges $e \in E$ and $w_z$ for vertices $z \in V$. According to Vizing's theorem[33] the chromatic index of every graph with maximal vertex-degree $d$ is either $d$ or $d+1$. Therefore there is a coloring of the edges $\mathcal{C} = \{C_i\}_i$ with $|\mathcal{C}| \leq d+1$. Denote $w(C_i) = \sum_{e \in C_i} w_e$ - the sum of the weights of all edges in $C_i$. Let $i_{max}$ be the "heaviest" color, i.e. the color with the property:

$$i_{max} = \arg\max_i w(C_i)$$

The heaviest color is at least as heavy as the average:

$$w(C_{i_{max}}) \geq \frac{1}{|\mathcal{C}|} \sum_i w(C_i) \geq \frac{1}{d+1} \sum_i w(C_i) \tag{5}$$

---
[9]If $j' \in V(E_\ell^j)$ then there exists an edge $e \ni j, j'$ so that $w_e > 0$. By Lemma 2.4 either $j' = j$ or $j' \in \text{Dep}^+(j)$.

And so:
$$(d+1) \sum_{e \in C_{i_{max}}} w_e = (d+1)w(C_{i_{max}}) \geq \sum_i w(C_i) = \sum_{e \in E} w_e \quad (6)$$

As a color, $C_{i_{max}}$ is a set of edges without common vertices, and can be seen as partition to disjoint pairs of some subset of $V$. Let $\mathcal{Q}$ be the partition of $[m]$ that we get by pairing all vertices not in $\bigcup_{e \in C_{i_{max}}} e$ in some way, and adding it to $C_{i_{max}}$. $\mathcal{Q}$ now satisfies:

$$\sum_\ell v(\mathcal{Q}_\ell) \geq [\sum_{z \in V} w_z + w(C_{i_{max}})] \geq \sum_{z \in V} w_z + \frac{1}{d+1} \sum_{e \in E} w_e$$
$$\geq \frac{1}{d+1}[sum_{z \in V} w_z + \sum_{e \in E} w_e] = \frac{v([m])}{d+1} \quad (7)$$

Given a partition $\mathcal{Q}$, let $h_\mathcal{Q}$ be the function:

$$h_\mathcal{Q}(X) = \frac{v(X)}{|X|\beta} \sum_\ell |\mathcal{Q}_\ell| = \frac{v(X)}{\beta}$$

Like in the proof of Lemma 3.5, we iteratively define a sequence of sets $S_i$ in the following way. Let $S_1 = X$. if there exists a set $T_1$ which satisfies $v(T_1) < h_\mathcal{Q}(T_1)$, assume w.l.o.g that $T_1$ is a union of sets from $\mathcal{Q}$ and define for every $i > 1$, $S_i = S_{i-1} \setminus T_{i-1}$. Because $T_i$ is a union of elements from $\mathcal{Q}$, so is $S_i$, and so $\mathcal{Q}$ induces a partition $\mathcal{Q}_{S_i}$ on $S_i$ and a $d+1$-CH function $h_{\mathcal{Q}_{S_i}}(T) = \frac{v(X)}{|S_i|\beta} \sum_{Q_\ell \in \mathcal{Q}_{S_i}: Q_\ell \subseteq T} |Q_\ell|$. If for some $i$ it holds that $h_{\mathcal{Q}_{S_i}}(T) \leq v(T)$ for all $T$, then $h_{\mathcal{Q}_{S_i}}(T)$ pointwise $\beta$-approximates $v$. Otherwise, the iterative process terminates at some $i_{max}$ because $|S_i|$ decreases every iteration. If the process terminates and none of the functions $h_{\mathcal{Q}_{S_i}}$ $\beta$-approximates $v$ at $X$, then we have two partitions of the set $X$: $\{\mathcal{Q}_\ell\}_\ell$ and $\{T_i\}_i$, so that every $\mathcal{Q}_\ell$ is a subset of some $T_j$. Therefore:

$$\frac{v(X)}{d+1} \leq \sum_\ell v(\mathcal{Q}_\ell) \leq \sum_i v(T_i) < \sum h_{\mathcal{Q}_{S_i}}(T_i) = \frac{v(X)}{\beta} \sum \frac{|T_i|}{|S_i|} \quad (8)$$

Where the first inequality is (7), the second is by super-modularity of the class PH-2, and third inequality is by construction. Rearranging terms yields:

$$\beta < (d+1) \sum \frac{|T_i|}{|S_i|}$$

Using equation (3) from the proof of lemma 3.5, we get:

$$\beta < (d+1) \sum \frac{|T_i|}{|S_i|} \leq (d+1) \sum_{k=1}^{\frac{m}{2}} \frac{1}{k} \leq (d+1)H_{m/2}$$

So for every $\beta \geq (d+1)H_{m/2}$, there is a 2-CH function that $\beta$-approximates $v$ at $X$.
□

## 3.2 Lower bounds

Proposition 3.9 shows a lower bound of $d$, which holds even for the more restricted class PH-2 ∩ SM-$d$, and even with respect to the best equilibrium.

**Proposition 3.9** *There exists an instance with one bidder with a SM-$d$ ∩ PH-2 valuation and one bidder that is interested in a single item, for which the price of stability of the single-bid auction is $d - \epsilon$ for every $\epsilon > 0$.*

*proof.* Consider an instance as described in the beginning of subsection 2.2, but with $d+1$ items. By adding $m-d-1$ items that have no value to any of the agents, the result follows directly.
□

In [11], a lower bound of $\Omega(\frac{\log m}{\log \log m})$ has been shown for the price of stability (PoS) of the single-bid auction with additive valuations. This bound carries over to valuations in PH-2 ∩ SM-$d$ for every $d$ (since additive valuations are a strict subclass of PH-2 ∩ SM-$d$). We conclude that the PoS for PH-2 ∩ SM-$d$ valuations is at least $\max(d, \Omega(\frac{\log m}{\log \log m}))$.

We show another example that simultaneously captures the two lower bounds above, i.e., an instance where agents have PH-2 ∩ SM-$d$ valuations, for which the PoS of the single bid auction is $\Omega(d + \frac{\log m}{\log \log m})$.

**Proposition 3.10** *If all agents have valuations in PH-2 ∩ SM-d, the PoS of the single bid auction w.r.t. pure Nash equilibria is at least $\Omega(d + \frac{\log m}{\log \log m})$.*

*proof.* Let $k$ be some number divisible by $d$, and let $[m]$ be composed of $k$ bundles-$\{B_0, ..., B_{k-1}\}$, where bundle $B_t$ is of size $|B_t| = k^t$. Let there be $4k + 1$ bidders. The first bidder (which we refer to as the "strong" bidder), has a valuation $w$ (which is PH-2 ∩ SM-$(d-1)$) as follows: The items in each bundle $B_t$ are divided to subsets of size $d$, and each of these groups is a $d$-star-graph in $w$'s hypergraph representation, with edge weight of $\frac{d}{d-1}k^{k-t}$. In total, $\forall t, w(B_t) = k^k$, and $w([m]) = k^{k+1}$. The next $2k$ bidders, with valuations marked $x_0, x'_0, ..., x_{k-1}, x'_{k-1}$ are as follows. First, denote $\lambda = \frac{d}{d+k}$. For each $t = 0, ..., (k-1)$, $x_t$ is additive, is only interested in $(1-\lambda) = \frac{k}{d+k}$ of the stars inside $B_t$, and only in the center of each star. For each center $j$ of any of these stars, $x_t(j) = k^{k-t-1}$. For all other items $j$, $x_t(j) = 0$. In addition $x'_t = x_t$. Note that the maximal value that bidder $x_t$ can get (by winning all of her desired items) is $x_t([m]) = x_t(B_t) = \frac{1}{d}(1-\lambda)k^{k-1}$. The final $2k$ bidders, with valuations marked $v_0, v'_0..., v_{k-1}, v'_{k-1}$ are as follows: for each $t = 0, ..., (k-1)$, $v_t$ is additive, is only interested in $\lambda = \frac{d}{d+k}$ of the stars inside $B_t$ (the stars that $x_t$ is not interested in), and only in the center of each star. For these special items, $v_t = k^{k-t} + \epsilon$. For all other items, $v_t = 0$. in addition $v'_t = v_t$. Note that if bidder $v_t$ wins all of her desired items the maximum value she can get is $v_t([m]) = v_t(B_t) = \frac{1}{d}\lambda k^k + \frac{1}{d}\lambda k^t \epsilon$.

The optimal allocation gives all items to the strong bidder and yields a social welfare of $k^{k+1}$. Due to best-response dynamics, in an equilibrium, every bidder $v_t, v'_t$ will bid exactly $k^{k-t} + \epsilon$ and every bidder $x_t, x'_t$ will bid exactly $k^{k-t-1}$. The special bidder will bid some number $b$. Whatever the value of $b$ is, she will win no more than two bundles, and no more than a fraction of $(1-\lambda) = \frac{k}{d+k}$ out of each of those two bundles. Assuming, w.l.o.g, that bidders $v_t$ and $x_t$ win every tie breaking, each of them wins all of her desired items. The social welfare will be:

$$\text{SW}(EQ) \leq 2(1-\lambda)k^k + k \cdot \frac{1}{d}\lambda k^k + \frac{1}{d}(1-\lambda)k^{k-1} = 2\frac{k}{d+k}k^k +$$

$$\frac{1}{d}\frac{d}{d+k}k^{k+1} + \frac{1}{d(d+k)}k^k = O(\frac{1}{d+k}k^{k+1})$$

This yields $PoS = \frac{\text{SW}(OPT)}{\text{SW}(EQ)} = \Omega(d+k) = \Omega(d + \frac{\log m}{\log \log m})$
□

## 4 Hybrid Single Bid Mechanisms

In this section we give a bound on the price of anarchy for general valuations by randomizing between the single bid auction and the *grand bundle auction*, described as follows.

**The Grand bundle auction** The grand-bundle auction solicits a single bid $b_i \in \mathbb{R}_+$ from each agent $i$, approaches the agents in decreasing order of their bids, and offers each agent $i$ the grand bundle $[m]$ for the price $b_i$, once an agent acquires $[m]$ the auction ends. Since the grand bundle auction solicits a single real-valued bid from each bidder, then runs a truthful mechanism, it is also an ALI mechanism.

We show that randomizing between the single bid auction and the grand bundle auction (by soliciting independent bids for the two auctions) gives price of anarchy of at most $O(\sqrt{m})$ for general valuations (Theorem 4.2).

Our proof proceeds as follows. We partition the space of valuation profiles into two disjoint subspaces, depending on whether there exists an optimal allocation where a single agent contributes at least $\frac{1}{\sqrt{m}}$ of OPT, or not. For the first subspace, we show that the grand bundle auction is $(\frac{1}{2\sqrt{m}}(1-e^{-1}), 1)$-smooth; for the second subspace we show that the single bid auction is $(\frac{1}{2\sqrt{m}} \cdot (1 - e^{-\sqrt{m}}), 1)$-smooth. This immediately implies that running both auctions independently and taking the better equilibrium gives price of anarchy of at most $O(\sqrt{m})$. However, a priori, it is not clear which auction to run. A naive approach would be to run both auctions simultaneously with independent bids, and apply the outcome of the better mechanism. As previously observed by [24], such an approach fails badly, as demonstrated by the following example:

**Example** Consider $m = n \geq 2$ agents. Each agent $i$ has value 1 for a unique item $a_i$, and a value of $1+\epsilon$ for the grand bundle $[m]$. If each agent submits a bid of 0 to the single-bid auction and a bid of $1+\epsilon$ to the grand bundle auction, the social welfare achieved by the grand bundle auction is $1+\epsilon$ by handing $[m]$ so some agent. This is a pure Nash equilibrium since no bidder can achieve non-negative utility in the grand bundle auction by deviating, and any unilateral deviation in the single bid auction that achieves non-negative utility for an agent, induces an outcome with social welfare at most 1, therefore the grand bundle auction will still be selected. Clearly the optimal allocation hands each agent $i$ its respective item of interest $a_i$, resulting in social welfare of $m$. This shows a price of anarchy of $m - \epsilon$ for every $\epsilon > 0$.

In [24] it is shown that the problem can be circumvented by introducing randomization. In this section we formalize this idea by using the smoothness framework.

**Definition 9** (Hybrid mechanism) *Given two mechanisms $\mathcal{M}$ and $\mathcal{M}'$, and a real number $0 < p < 1$, the hybrid mechanism $(\mathcal{M}, \mathcal{M}', p)$ solicits from each agent $i$ two actions, $a_i, a_i'$, and runs $\mathcal{M}(\mathbf{a})$ with probability $p$ and $\mathcal{M}'(\mathbf{a}')$ with probability $1-p$.*

It follows by definition that a hybrid mechanism that is composed of two ALI mechanisms is also an ALI mechanism.

The following lemma establishes the smoothness of a hybrid mechanism.

**Lemma 4.1** *Let $\mathcal{V}$ and $\mathcal{V}'$ be spaces of valuation profiles. Suppose mechanism $\mathcal{M}$ is $(\lambda, \mu)$-smooth w.r.t. valuation profiles in $\mathcal{V}$, and mechanism $\mathcal{M}'$ is $(\lambda', \mu')$-smooth w.r.t. valuation profiles in $\mathcal{V}'$. Then, for every $p$, the hybrid mechanism $(\mathcal{M}, \mathcal{M}', p)$ is $(p \cdot \lambda, \max\{\mu, 1\})$-smooth w.r.t. valuation profiles in $\mathcal{V}$ and $((1-p) \cdot \lambda', \max\{\mu', 1\})$-smooth w.r.t. valuation profiles in $\mathcal{V}'$.*

*proof.* Consider a valuation profile $\mathbf{v} \in \mathcal{V}$. Consider an arbitrary action profile $(\mathbf{a}, \mathbf{a}')$, where $\mathbf{a} = (a_1, \ldots, a_n)$ and $\mathbf{a}' = (a_1', \ldots, a_n')$. Let $P_i$ and $P_i'$ denote the payments of mechanisms $\mathcal{M}$ and $\mathcal{M}'$ respectively, and similarly for utilities and values. Utilities $(u_i^p)$, values $(v_i^p)$, and payments $(P_i^p)$, denote the expected value of those quantities for agent $i$ in the hybrid mechanism $(\mathcal{M}, \mathcal{M}', p)$ (e.g. for payments, $P_i^p(\mathbf{a}, \mathbf{a}') = p \cdot P_i(\mathbf{a}) + (1-p) \cdot P_i'(\mathbf{a}')$). Let $a_i^*(\mathbf{v})$ be the deviation given by the smoothness of mechanism $\mathcal{M}$. For ease of exposition denote $a_i^*(\mathbf{v})$ by $a_i^*$ and assume that $a_i^*$ is a pure strategy. By considering the utility of each agent $i$ at the action profile $((a_i^*, \mathbf{a}_{-i}), \mathbf{a}')$ and then using the linearity of expectation:

$$\sum_i u_i^p((a_i^*, \mathbf{a}_{-i}), \mathbf{a}') = \sum_i p \cdot u_i(a_i^*, \mathbf{a}_{-i}) + (1-p) u_i'(\mathbf{a}') = p \sum_i u_i(a_i^*, \mathbf{a}_{-i}) + (1-p) \sum_i u_i'(\mathbf{a}')$$

By smoothness of $\mathcal{M}$ it holds that:

$$\sum_i u_i^p((a_i^*, \mathbf{a}_{-i}), \mathbf{a}') \geq p \cdot \left( \lambda \cdot \mathrm{SW}(OPT(\mathbf{v})) - \mu \cdot \sum_i P_i(\mathbf{a}) \right) + (1-p) \cdot \sum_i u_i'(\mathbf{a}')$$

$$= p \cdot \lambda \cdot \mathrm{SW}(OPT(\mathbf{v})) - \mu \cdot p \cdot \sum_i P_i(\mathbf{a}) + (1-p) \cdot \sum_i (v_i'(\mathbf{a}') - P_i'(\mathbf{a}'))$$

By $v_i'(\mathbf{a}') \geq 0$ we get:

$$\sum_i u_i^p((a_i^*, \mathbf{a}_{-i}), \mathbf{a}') \geq p \cdot \lambda \cdot \mathrm{SW}(OPT(\mathbf{v})) - \max\{\mu, 1\} \sum_i (p \cdot P_i(\mathbf{a}) + (1-p) \cdot P_i'(\mathbf{a}'))$$

$$= p \cdot \lambda \cdot \mathrm{SW}(OPT(\mathbf{v})) - \max\{\mu, 1\} \sum_i P_i^p(\mathbf{a}, \mathbf{a}'),$$

where the last equality follows by the definition of the hybrid mechanism. Symmetrically, for every valuation profile $\mathbf{v}' \in \mathcal{V}'$ the mechanism is $((1-p) \cdot \lambda', \max\{\mu', 1\})$-smooth with respect to valuations in $\mathcal{V}'$.
□

We now establish our main theorem. While each of the grand bundle and single bid auctions has a price of anarchy of $\Omega(m)$ for general valuations, the hybrid mechanism gives price of anarchy of at most $O(\sqrt{m})$.

**Theorem 4.2** *The hybrid mechanism composed of the single-bid and the grand-bundle auctions with $p = 1/2$ is $(\frac{1}{4\sqrt{m}}(1-e^{-1}), 1)$-smooth for general valuations.*

We first consider valuation profiles in which the optimal welfare can be approximated by allocating the grand bundle.

**Lemma 4.3** *If for a class of valuation profiles $\mathcal{V}$, for every $\mathbf{v} \in \mathcal{V}$ there exists an agent $i^*$ so that $v_{i^*}([m]) \geq \beta \cdot SW(OPT(\mathbf{v}))$, then the grand-bundle auction is a $(\beta \cdot (1-e^{-1}), 1)$-smooth mechanism.*

*proof.* Consider a valuation profile $\mathbf{v}$, and assume there exists an agent $i^*$ so that $v_{i^*}([m]) \geq \beta \cdot \mathrm{SW}(OPT(\mathbf{v}))$. Consider an arbitrary bid profile $\mathbf{b} = (b_1, \ldots, b_n)$, and let $b'(\mathbf{b})$ be the winning bid in $\mathbf{b}$. If agent $i^*$ deviates to a deterministic bid $t \leq v_{i^*}([m])$, then $i^*$ can acquire the grand bundle for sure only if $t > b'(\mathbf{b})$. Therefore:

$$u_{i^*}(t, \mathbf{b}_{-i^*}) \geq (v_{i^*}([m]) - t) \cdot \mathbf{1}\{t > b'(\mathbf{b})\}$$

Note that $\sum_{i \in N} P_i(\mathbf{b}) = b'(\mathbf{b})$. Consider the randomized deviation $B_{i^*}'$ distributed by the density function:

$$f(t) = \frac{1}{v_{i^*}([m]) - t}$$

on the support $[0, (1-e^{-1})v_{i^*}([m])]$. Then:

$$\mathbb{E}\left[u_{i^*}(B_{i^*}', \mathbf{b}_{-i^*})\right] \geq \int_{b'(\mathbf{b})}^{(1-e^{-1})v_{i^*}([m])} (v_{i^*}([m]) - t) f(t) \mathrm{d}t$$

$$= \int_{b'(\mathbf{b})}^{(1-e^{-1})v_{i^*}([m])} 1 \cdot \mathrm{d}t$$

$$= (1-e^{-1}) v_{i^*}([m]) - b'(\mathbf{b})$$

$$\geq \beta \cdot (1-e^{-1}) \cdot \mathrm{SW}(OPT(\mathbf{v})) - \sum_{i \in N} P_i(\mathbf{b})$$

Since all other agents can acquire a non-negative utility, we conclude.
□

We now consider valuation profiles in which the optimal welfare can be well-approximated by "small" allocations.

**Lemma 4.4** *If for every valuation profile $\mathbf{v}$ in a class of valuation profiles $\mathcal{V}$ there exists an allocation $S^*$ so that $SW(S^*) \geq \beta \cdot SW(OPT(\mathbf{v}))$ and $|S_i^*| \leq \gamma$ for every agent $i$, then for every $c > 0$ the single bid auction is $(c \cdot (1 - e^{-1/c})\beta, c \cdot \gamma)$-smooth w.r.t. $\mathcal{V}$.*

*proof.* Consider a valuation profile $\mathbf{v}$ and let $S^*$ be an allocation that $\beta$-approximates the optimal allocation $OPT(\mathbf{v})$. Consider an arbitrary bid profile $\mathbf{b} = (b_1, \ldots, b_n)$. Denote by $p_j(\mathbf{b})$ the price of item $j$ under bid profile $\mathbf{b}$. If agent $i$ deviates to a deterministic bid $t < \frac{v_i(S_i^*)}{|S_i^*|}$, she can acquire the set $S_i^*$ only if $t > \max_{j \in S_i^*} p_j(\mathbf{b})$. Therefore:

$$u_i(t, \mathbf{b}_{-i}) \geq (v_i(S_i^*) - t \cdot |S_i^*|) \cdot \mathbf{1}\{t > \max_{j \in S_i^*} p_j(\mathbf{b})\}$$

Given $\mathbf{v}$, and a bundle of items $B$, let $D_i(B)$ be $i$'s average value-per-item of the bundle $B$, i.e.,

$$D_i(B) = \frac{v_i(B)}{|B|}$$

Furthermore, for ease of notation let $D_i^* = D_i(S_i^*)$. Consider the randomized deviation $B_i'$ distributed by the density function:

$$f(t) = c \cdot \frac{1}{D_i^* - t}$$

on the support $\left[0, c \cdot (1 - e^{-1/c})D_i^*\right]$. Then:

$$\mathbb{E}\left[u_i(B_i', \mathbf{b}_{-i})\right] \geq \int_{\max_{j \in S_i^*} p_j(\mathbf{b})}^{c \cdot (1-e^{-1/c})D_i^*} (v_i(S_i^*) - t \cdot |S_i^*|) f(t) \mathrm{d}t$$

$$= c \cdot \int_{\max_{j \in S_i^*} p_j(\mathbf{b})}^{c \cdot (1-e^{-1/c})D_i^*} \frac{v_i(S_i^*) - t \cdot |S_i^*|}{D_i^* - t} \mathrm{d}t$$

$$= c \cdot \int_{\max_{j \in S_i^*} p_j(\mathbf{b})}^{c \cdot (1-e^{-1/c})D_i^*} \frac{|S_i^*|(D_i^* - t)}{D_i^* - t} \mathrm{d}t$$

$$= c \cdot (1 - e^{-1/c}) v_i(S_i^*) - c \cdot \max_{j \in S_i^*} p_j(\mathbf{b}) \cdot |S_i^*|$$

Summing over all agents we get:

$$\sum_i \mathbb{E}\left[u_i(B_i', \mathbf{b}_{-i})\right] \geq c \cdot (1 - e^{-1/c}) SW(S^*) - c \cdot \sum_i \max_{j \in S_i^*} p_j(\mathbf{b}) \cdot |S_i^*| \quad (9)$$

$$\geq c \cdot (1 - e^{-1/c}) SW(S^*) - c \cdot \sum_i |S_i^*| \sum_{j \in S_i^*} p_j(\mathbf{b}) \quad (10)$$

For every $i$ it holds that $|S_i^*| \leq \gamma$ therefore:

$$\sum_i \mathbb{E}\left[u_i(B_i', \mathbf{b}_{-i})\right] \geq c \cdot (1 - e^{-1/c}) SW(S^*) - c \cdot \gamma \sum_i \sum_{j \in S_i^*} p_j(\mathbf{b})$$

$$\geq c \cdot (1 - e^{-1/c}) \beta \cdot SW(OPT(\mathbf{v})) - c \cdot \gamma \sum_j p_j(\mathbf{b})$$

As required.
□

The above lemmas lead to the following definition.

**Definition 10** *A valuation profile* $\mathbf{v}$ *is $z$-lopsided if there exists an optimal allocation $S^*$ so that at least half of the social welfare is due to agents that were allocated a bundle with at least $z$ goods, i.e., if $\sum_{i \in A} v_i(S_i^*) \geq \frac{1}{2} SW(S^*)$, where $A \subseteq N$ and for every $i \in A$ it holds that $|S_i^*| \geq z$. We denote by $LOP(z)$ the class of all $z$-lopsided valuation profiles.*

The following lemma is implied by Lemma 4.3.

**Lemma 4.5** *The grand-bundle auction is a $(\frac{z}{2m} \cdot (1 - e^{-1}), 1)$-smooth mechanism with respect to valuation profiles in $LOP(z)$.*

*proof.* Fix a valuation profile $\mathbf{v} \in LOP(z)$. There exists an allocation $S^*$ and a set of agents $A \subseteq N$ so that $SW(S^*) = SW(OPT(\mathbf{v}))$ and for every $i \in A$ it holds that $|S_i^*| \geq z$, and that $\sum_{i \in A} v_i(S_i^*) \geq \frac{1}{2}SW(S^*)$. Since $|A| \cdot z \leq m$ it must be that $|A| \leq \frac{m}{z}$. Therefore, there must exist an agent $i^* \in A$ so that $v_{i^*}(S_{i^*}^*) \geq \frac{1}{|A|} \sum_{i \in A} v_i(S_i^*) \geq \frac{z}{2m} SW(S^*)$. The assertion of the lemma is established by applying lemma 4.3.
□

Similarly, the following lemma is implied by Lemma 4.4.

**Lemma 4.6** *For every $c > 0$, the single-bid auction is a $(\frac{c}{2} \cdot (1 - e^{-1/c}), c \cdot z)$-smooth mechanism with respect to valuation profiles $\mathbf{v} \notin LOP(z)$.*

*proof.* Fix a valuation profile $\mathbf{v} \notin LOP(z)$. Consider an optimal allocation $S^*$. Consider the set of agents $A = \{i \in N : |S_i^*| < z\}$. Since $\mathbf{v} \notin LOP(z)$ it must be that $\sum_{i \in A} v_i(S_i^*) > \frac{1}{2}SW(S^*)$, otherwise the set of agents $N \setminus A$ would imply that $\mathbf{v} \in LOP(z)$. Therefore, by lemma 4.4, for every $c > 0$ the single bid auction is $(\frac{c}{2} \cdot (1 - e^{-1/c}), c \cdot z)$-smooth with respect to valuation profiles not in $LOP(z)$.
□

To conclude the proof of Theorem 4.2, note that Lemma 4.5 implies that the grand-bundle auction is $(\frac{1}{2\sqrt{m}}(1 - e^{-1}), 1)$-smooth w.r.t. valuation profiles in $LOP(\sqrt{m})$, and Lemma 4.6, with $c = \frac{1}{\sqrt{m}}$, implies that the single-bid auction is $(\frac{1}{2\sqrt{m}} \cdot (1 - e^{-\sqrt{m}}), 1)$-smooth w.r.t. valuations not in $LOP(\sqrt{m})$. It follows by Lemma 4.1 that the hybrid mechanism is $(\frac{1}{4\sqrt{m}}(1 - e^{-1}), 1)$-smooth, as desired.

## 4.1 Lower bounds

This guarantee is tight up to a constant of $\frac{1}{4}(1 - e^{-1})$. we show a lower bound of $\sqrt{m}$ for the PoA of the hybrid mechanism in every PNE.

**Proposition 4.7** *There exists a valuation profile $\mathbf{v}$ for which the PoA of the Hyb mechanism with regard to pure Nash equilibria is at least $\sqrt{m}$.*

*proof.* For some $k$, Consider $2k$ bidders with valuation functions $v_t, x_t$ for $t = 1, ..., k$ and items $1, 2, \ldots, k^2 = m$. In a slight abuse of notation we will say "bidder $v_t$" and mean "the bidder with valuation $v_t$". Divide $[m]$ into $k$ bundles of size $k$ each - $B_1, ..., B_k$ with $B_t = \{(t - 1) \cdot k + j : j = 1, \ldots, k\}$. For every $t = 1, ..., k$, $v_t \in$ PH-2 $\cap$ SM-$(k-1)$, has a star shaped valuation where the vertex set of the star is $B_t$, the center is item $(t - 1)k + 1$, and the weight of each edge is 1. Also, for very $t$, bidder $x_t$ is interested only in the item $(t - 1)k + 1$ with a value of $\frac{k-1}{k} + \epsilon$. Let $\mathbf{b}$ denote a PNE of SBA and $\mathbf{b}'$ a PNE of GB auction. The profile $(\mathbf{b}, \mathbf{b}')$ is a pure Nash equilibrium of Hyb. Clearly the optimal allocation gives each bundle $B_t$ to bidder $v_t$, yielding a social welfare of $k(k - 1)$. By the same argument that is used to show Observation 2.3, if the SB mechanism is played than bidders $v_t$ win nothing and bidders $x_t$ win all star centers (items of the form $(t - 1)k + 1$ for $t = 1, 2 \ldots, k$) and get a total social welfare of $k(\frac{k-1}{k} + \epsilon) = k - 1 + k\epsilon$. The total value of each bidder for the grand bundle $[m]$ is at most $k - 1$ so the GB auction cannot achieve a social welfare of more than $k - 1$. We get that if Hyb is played, regardless of which of the two mechanisms (SB or GB) is actually played, the obtained social welfare is no more than $k - 1 + k\epsilon$, which is arbitrarily close to $\frac{1}{k} = \frac{1}{\sqrt{m}}$ of the optimal.
□

## 5  Discussion

In this paper we study simple mechanisms for settings that exhibit complementarities. We focus on the single bid auction, which has been shown to have a logarithmic PoA for complement free valuations. We show upper and lower bounds on the PoA when agents have complementarities, captured by the MPS-$d$ hierarchy. We also show that randomizing between the single bid auction and the grand bundle auction gives PoA of $O(\sqrt{m})$ for general valuations. Our results leave a gap between the lower and upper bounds on the PoA of the single bid auction when applied to MPS-d valuations. In the full version we show that our upper bound is essentially tight with respect to our proof technique. In particular, we show that the pointwise approximation of MPS-d by $(d+1)$-CH valuations is tight (up to a $\log m$ factor). A major open problem is to find the price of anarchy of the single bid auction for SM-$d$ valuations.

## 6  Acknowledgments

We thank Michael Krivelevich and Asaf Shapira for fruitful discussions.

## A  Limitations on the pointwise approximation method for PS-$d$

In this section we discuss the limitations of the pointwise approximation method for valuations in PS-$d$. It remains an interesting open question - what is the real approximation ratio between $(d+1)$-CH and PS-$d$, and how does it depend on the number of items $m$? We show progress in answering this question by proving various lower bounds for the approximation ratio of PS-$d$ by the classes $k$-CH for all $k \leq d+1$. The following proposition shows that using $k$-CH valuations where $k < d+1$ cannot improve our results.

**Proposition A.1** *For all d, and all $k < d + 1$, there exists a valuation $v \in$ PS-d such that if $v' \in k$-CH pointwise $\beta$-approximates $v$ at $[m]$, then $\beta \geq \binom{d}{k-1}$.*

*proof.* We show that there exists a valuation $v \in$ PS-$d$ such that for all $v' \in k$-CH, it holds that $\frac{v([m])}{v'([m])} \geq \binom{d}{k-1}$. Set $m = d+1$ and consider the valuation $v \in$ PS-$d$ with the hypergraph representation that contains all of the possible hyper-edges of size $k$, and gives each hyper-edge a weight of 1. There are $\binom{d+1}{k}$ such hyper-edges and therefore $v([m]) \leq \binom{d+1}{k}$. Assume $v' \in k$-CH and that $v'$ $\beta$-approximates $v$ at $[m]$. Because $v' \in k$-CH, all edges that are assigned positive weight by $v'$ must be disjoint. Therefore $v'$ cannot assign positive weight to more than $\frac{d+1}{k}$ hyper edges of size $k$. Furthermore, by the definition of $\beta$-approximation, for every $T \subseteq [m]$ it holds that $v'(T) \leq v(T)$. Specifically for all hyper edges $e$ with $|e| < k$, $v'(e) \leq v(e) = 0$, and for all hyper edges $e$ with $|e| = k$, $v'(e) \leq v(e) = 1$. Therefore, $v'([m]) \leq \frac{d+1}{k}$. In total we get:

$$\beta \geq \frac{v([m])}{v'([m])} \geq \frac{k}{d+1} \binom{d+1}{k} = \binom{d}{k-1} \tag{11}$$

□

Next, we show two lower bounds on the approximation ratio of PS-d by the class $(d+1)$-CH. The following is from [9].

**Theorem A.2** *For $d = 2, 3, 5, 7$ and $d \geq 10$, there exist d-regular graphs, for which the shortest cycle is of length larger than $\log_d(m/4)$.*

**Proposition A.3** *For $d = 2, 3, 5, 7$ and $d \geq 10$, there exists a large enough m and a valuation $v \in$ PS-d, such that if $v' \in (d+1)$-CH and $v'$ pointwise $\beta$-approximates $v$ at $[m]$ then $\beta \geq d$.*

*proof.* Consider the valuation $v$ with the hypergraph representation given by having a weight 1 on each edge from the graph given by theorem A.2. Since the graph is $d$ regular, there are $d \cdot m$ edges, therefore $v([m]) = d \cdot m$. For large enough $m$, the shortest cycle in the graph is larger than $d+1$, therefore in any set of at most $k \leq d+1$ nodes, there will be at most $k-1$ edges connecting two nodes from the set. Let $v'$ be a $(d+1)$-CH valuation that $\beta$-approximates $v$. By definition of pointwise $\beta$-approximation, for every item $j$ it holds that $v'(\{j\}) \leq v(\{j\}) = 0$ for every edge $e$ it holds that $v'(e) \leq v(e) = 1$. Let $Q_1, \ldots Q_\ell$ be the sets that form the valuation $v'$. For any of the sets $Q_i$, it must hold that $v'(Q_i) \leq |Q_i|$ therefore $v'([m]) \leq m$. As a result $\frac{v([m])}{v'([m])} \geq d$ which implies $\beta \geq d$. □

Note that the requirement $\log_d(m/4) \geq d+1$ translates to $m = \Omega(d^d)$. The next result is a slightly less tight lower bound, but for a more general case.

**Proposition A.4** *For large enough $d$, and $m \geq d^2$, there exists a valuation $v \in$ PS-$d$, such that if $v' \in (d+1)$-CH and $v'$ pointwise $\beta$-approximates $v$ at $[m]$ then $\beta = \Omega(\frac{d}{\log d})$.*

For the proof of proposition A.4, we will use random graphs to show there exists a valuation $f \in$ PS-$d$ such that for every $g \in (d+1)$-CH, $\frac{f([m])}{g([m])} \geq C \cdot \frac{d}{\log d}$ for some constant $C$. Let $G = (V, E)$ be a graph, and denote $e(S) = |\{e = ij \in E \text{ such that } i, j \in S\}|$ (the number of edges in $G$ with both endpoints in $S$). For the proof of proposition A.4 we use the following lemma:

**Lemma A.5** *For large enough $d$, there exists a graph $G = (V, E)$ on $m = d^2$ vertices which satisfies:*

1. *Every vertex set $S \subseteq V$ with $|S| = k \leq d+1$ satisfies $e(S) \leq 12k \log d$.*

2. *The maximal vertex degree $\Delta(G)$ satisfies $\Delta(G) \leq d$*

3. *$|E| \geq \frac{1}{9}d^3$*

Using the graph $G$ from lemma A.5 we prove proposition A.4:

*proof.* [of proposition A.4 ] Assume that $d$ is large enough for $G = (V, E)$ from lemma A.5 to exist, and assume $d \leq \sqrt{m}$. Let $f$ be a graphical valuation on $[m]$, constructed in the following way: divide $[m]$ into $T = \frac{m}{d^2}$ bundles of size $d^2$ each $= B_1, B_2, ..., B_T$. For each $B_t$, fix some bijection $\pi_t : B_t \to V$ and let the edges in $B_t$ correspond to edges in $G$ as induced by $\pi_t$. Let each edge in $B_t$ have a weight of 1, and each vertex - a weight of 0. Thus, for all $t$, $f(B_t) = \Omega(d^3)$ and $f([m]) = \Omega(d^3 \frac{m}{d^2}) = \Omega(m \cdot d)$, furthermore - $f \in$ PS-$d$.

Now, consider any $d+1$-CH valuation function $g$ on $[m]$. Denote $\mathbf{Q}^g = \{Q_i^g\}_{i \in \mathcal{I}(g)}$ the supporting item sets for $g$. By definition $|Q_i^g| \leq d+1$ for all $i \in \mathcal{I}(g)$. Assume that $g$ satisfies $g(S) \leq f(S)$ for all $S \subseteq [m]$. To finish it is enough to prove that $g([m]) = O(m \log d)$. $g \leq f$, and by the construction of $f$ we get that for any item set $Q_i^g$:

$$\hat{v}_g \cdot |Q_i^g| = g(Q_i^g) \leq f(Q_i^g) = 1 \cdot e(Q_i^g)$$
$$= \sum_t e(B_t \cap Q_i^g) \leq \sum_t 12 |B_t \cap Q_i^g| \cdot \log d = 12 |Q_i^g| \cdot \log d$$

We get that $\hat{v}_g \leq 12 \log d$. So for $g([m])$ we get:

$$g([m]) = \hat{v}_g \cdot \sum_{i \in \mathcal{I}(g)} |Q_i^g| \leq \hat{v}_g \cdot m \leq 12 \cdot m \log d$$

as required. □

We now turn to prove lemma A.5.

*proof.* [Of Lemma A.5] Consider a random graph $G(m, p)$ with $m = d^2$ vertices and $p = \frac{1}{2d}$ the independent probability for the existence of each edge. We will show that for large enough $d$, with positive probability $G(m, p)$ satisfies all three requirement simultaneously and therefore such a graph must exist. For this it is enough to show that each of the requirements by itself is fulfilled with high probability (abbreviated w.h.p.), i.e. the probability that the requirement is fulfilled tends to 1 as $d$ increases.

1. For $S$ with $|S| = k \leq \log d$ the claim is trivial, there are at most $\frac{1}{2}k^2$ edges in $S$, and if $k \leq \log d$ then $\frac{1}{2}k^2 \leq k \log d$. For $k > \log d$, the number of edges in any set $S$ of size $k \leq d+1$ is a binomial random variable $X_S = Bin(\binom{k}{2}, \frac{1}{2d})$. Its expectation:

$$\mu = \mathbb{E}[X_S] = \binom{k}{2}\frac{1}{2d} = \frac{k(k-1)}{4d}$$

Using a Chernoff bound we get (for $\epsilon > 1$):

$$Pr\{X_S \geq (1+\epsilon)\mu\} \leq \exp(-\frac{\epsilon^2}{2+\epsilon}\mu) \leq \exp(-\frac{1}{2}\epsilon\frac{k^2}{4d}) = \exp(-\frac{k^2}{8d}\epsilon)$$

There are $\binom{d^2}{k}$ vertex subsets of size $k$. A standard bound for $\binom{n}{k}$ is:

$$\binom{d^2}{k} \leq \left(\frac{ed^2}{k}\right)^k$$

Let $\epsilon = C\frac{d}{k}\log(\frac{d^2}{k})$ for some $C$ large enough to be determined later. Note that:

$$(1+\epsilon)\mu = [1 + C\frac{d}{k}\log(\frac{d^2}{k})]\frac{k(k-1)}{4d} = \frac{1}{2}C(k-1)\log d - \frac{1}{4}C(k-1)\log k + \frac{k(k-1)}{4d} \leq \frac{1}{2}Ck\log d$$

Let $Y_S$ be the indicator variable that is equal to 1 if $X_S \geq \frac{1}{2}Ck\log d \geq (1+\epsilon)\mu$ and 0 otherwise. Denote

$$Y_k = \sum_{S \subseteq V, |S|=k} Y_S$$

Using the union bound we get:

$$\mathbb{E}[Y_k] = \mathbb{E}[\sum_{S \subseteq V, |S|=k} Y_S] = \sum_{S \subseteq V, |S|=k} \mathbb{E}[Y_S] = \sum_{S \subseteq V, |S|=k} Pr\{Y_S = 1\}$$

$$\leq (\frac{ed^2}{k})^k \cdot e^{-\frac{k^2}{8d}\epsilon} \leq \exp(k(\log(\frac{d^2}{k})+1) - \frac{C}{8}k\log(\frac{d^2}{k}))$$

$$\leq \exp(-(\frac{C}{8}-2)k\log(\frac{d^2}{k})) \leq \exp(-k\log(\frac{d^2}{k}))$$

For all of the inequalities in the above calculation to hold it's enough to take $C > 24$. We see that the expected number of sets of size $k$ with more than $12k\log d$ edges is vanishingly small. Thus, Markov's inequality implies:

$$Pr\{Y_k \geq 1\} \leq \mathbb{E}[Y_k] \leq e^{-k\log(\frac{d^2}{k})}$$

Again using the union bound we get:

$$Pr\{\exists 1 \leq k \leq (d+1) : Y_k \geq 1\} \leq \sum_{k=1}^{d+1} e^{-k\log(\frac{d^2}{k})}$$

$$\leq (d+1)e^{-2\log d} = \frac{d+1}{d^2}$$

so w.h.p every vertex subset $S \subseteq V$ with $|S| = k \leq d+1$ satisfies $e(S) \leq 12k\log d$.

2. The degree $deg(x)$ of each vertex $x \in V$ is a binomial random variable $B(m-1,p) = B(d^2-1, \frac{1}{2d})$. Its expectation is $\mathbb{E}[deg(x)] = \frac{d^2-1}{2d} = \frac{1}{2}d - \frac{1}{2d}$. Using Chernoff again:

$$Pr\{deg(x) > d\} \leq e^{-\frac{1}{6}d}$$

Using the union bound again:

$$Pr\{\Delta(G) > d\} \leq \sum_{x \in V} \Pr\{deg(x) > d\} \leq d^2 e^{-\frac{1}{6}d}$$

so w.h.p. $\Delta(G) \leq d$.

3. The total number of edges in $G$ is a binomial random variable $B(\binom{d^2}{2}, \frac{1}{2d})$ with expectation $\mathbb{E}[|E|] = \frac{1}{4}d^3 - \frac{1}{4}d$. With Chernoff we get:

$$Pr\{|E| \leq \frac{1}{8}d(d^2-1)\} \leq e^{-\frac{1}{32}d^3 + \frac{1}{32}d}$$

so w.h.p. $|E| \geq \frac{1}{8}d^3 - \frac{1}{8}d \geq \frac{1}{9}d^3$.

□

Next, we show that our analysis of the greedy algorithm in the proof of lemma 3.5 is almost tight:

**Proposition A.6** *If $d < \sqrt{m}$, there exists a valuation $v \in$ PS-$d$, for which the partition $\{\mathcal{Q}_\ell\}_\ell$ given by algorithm 1 satisfies:*

$$v([m]) = d \sum_\ell v(\mathcal{Q}_\ell)$$

This shows the analysis of algorithm 1 is almost tight because for the partition that is returned by the algorithm we show that: $v([m]) \leq (d+2) \sum_\ell v(\mathcal{Q}_\ell)$.

*proof.* Let $G = (V, E)$ be a graph with vertices that correspond to items in the auction, i.e. $V = [m]$, constructed in the following way: divide $V$ to $T = \lfloor \frac{m}{d^2+1} \rfloor$ bundles of size $d^2 + 1$ each - $B_1, ... B_T$. Number all of the items in $\bigcup_t B_t$ by ordered pairs - $(t, j) \in \{1, ..., T\} \times \{0, 1, ..., d^2\}$ such that $B_t = \{(k, j) : k = t\}$, i.e. the first coordinate is the bundle number for the item and the second coordinate is the number inside the bundle. The set of edges $E$ is defined in the following way:

$$E_t^{center} = \left\{ \{(t, d^2), (t, kd)\} : k = 0, ..., (d-1) \right\}$$

$$E_t^{rim} = \left\{ \{(t, kd), (t, kd+j)\} : k = 0, ..., (d-1), j = 1, ..., (d-1) \right\}$$

$$E = \bigcup_{t=1,...,T} E_t = \bigcup_{t=1,...,T} (E_t^{center} \cup E_t^{rim})$$

Note that $E_t$ is the set of edges in $G$ with both ends in $B_t$, and there are no edges $e = (x, y) \in E$ with $x \in B_{t_1}$ and $y \in B_{t_2}$, i.e. there are no crossing edges between different bundles. The valuation $v$ is described, as usual, via its graphical representation - it gives a weight of 0 to each individual item, a weight of $\frac{1}{t}$ to edges $e \in E_t^{center}$ and a weight of $\frac{1}{t} - \epsilon$ (for an arbitrary small $\epsilon > 0$) to edges in $E_t^{rim}$.

First note that $v \in$ PS-$d$ because all edges have non-negative weight and no item has more than $d$ neighbors.

**Lemma A.7** *$v$ satisfies the following properties:*

1. The output of algorithm 1 when run on $v$ returns the partition:
$$\{Q_i\}_i = \{Q_t\}_{t=1,...,T} = \{(t, kd) : k = 0, ..., d\}_{t=1,...,T}$$

2. For every $t=1,...,T$:
$$\{e \in E : e \subseteq Q_t\} = E_t^{center}$$

Using Lemma A.7, we calculate:
$$v([m]) = \sum_{t=1,...,T} v(B_t) = \sum_{t=1,...,T} d\frac{1}{t} + d(d-1)(\frac{1}{t} - \epsilon)$$
$$= \sum_{t=1,...T} \left[d^2\frac{1}{t} - d(d-1)\epsilon\right] = -Td(d-1)\epsilon + d\sum_{t=1,...,T} d\frac{1}{t}$$
$$= -Td(d-1)\epsilon + d\sum_{t=1,...,T} v(Q_t)$$

and by choosing $\epsilon$ to be small enough this can be arbitrarily close to $d\sum_{t=1,...,T} v(Q_t)$ as required. □

*proof.* [Of Lemma A.7] We prove the properties of the lemma by running the algorithm on the input $v$. In the first iteration of step 3, the algorithm chooses

$$Q_1 \in \arg \max_{\substack{A \subseteq [m] \\ |A|=d+1}} \{v(A)\}$$

which is exactly the set $\{(1, kd) : k = 0, ..., d\}$ that contains in it all the edges in $E_1^{center}$, and has a weight of $d$. Note that all edges in $B_1$ have at least one endpoint in $Q_1$, thus adding the items in $B_1 \setminus Q_1$ to any future $Q_t$ will not add any value to it. In a similar way one can see that in the $t$-th iteration of step 3 the set that will be chosen as $Q_t$ will be $\{(t, kd) : k = 0, ..., d\}$, the edges strictly contained in it are exactly $E_t^{center}$ and it has a weight of exactly $\frac{d}{t}$. □

## B Omitted Proofs

### B.1 Omitted Part of the Proof of Lemma 3.5

If $|X|$ is not divisible by $d+1$, exactly one of the partition elements $Q_\ell$ is strictly smaller than $d+1$, and hence there is exactly one index $\hat{i}$ for which $T_{\hat{i}}$ is not a multiple of $d+1$. Denote $r = |T_{\hat{i}}| \mod (d+1)$ and $t_i = \frac{|T_i|}{d+1}$. Define $r_i = |S_i| \mod (d+1)$ and note that for all $i \leq \hat{i}$, $r_i = r$ and for all $i > \hat{i}$, $r_i = 0$. Finally, denote $s_i = \frac{|S_i|-r_i}{d+1} = \lfloor\frac{|S_i|}{d+1}\rfloor$ Now calculate:

$$\sum_i \frac{|T_i|}{|S_i|} = \sum_{i\neq\hat{i}} \frac{|T_i|}{|S_i|} + \frac{|T_{\hat{i}}|}{|S_{\hat{i}}|} \leq \sum_{i\neq\hat{i}} \frac{|T_i|}{|S_i|-r_i} + 1 = \sum_{i\neq\hat{i}} \frac{t_i}{s_i} + 1 \leq \sum_i \sum_{j=0}^{t_i-1} \frac{1}{s_i} + 1$$
$$\leq \sum_i \sum_{j=0}^{t_i-1} \frac{1}{s_i-j} + 1 = \sum_{j=0}^{s_1-1} \frac{1}{s_1-j} + 1 = H_{s_1} + 1 \leq H_{\lfloor\frac{|X|}{d+1}\rfloor} + 1$$

24